\documentclass[preprint,authoryear,12pt]{arxivarticle}

\usepackage{graphicx}

\usepackage{amssymb, amsmath}

\begin{document}

\begin{frontmatter}



\title{Mathematical Modeling of Dynamics for Partially Filled Shells of Revolution}


\author[IK]{I. Kononenko}
\author[OK]{O. Kononenko}

\address[IK]{Kharkiv National University, Kharkiv, Ukraine}
\address[OK]{CERN, Geneva, Switzerland}

\begin{abstract}
In this work we study the dynamic behaviour of compound shells of revolution partially filled with an ideal incompressible fluid based on boundary-value problems. New analytical mathematical model with corresponding discrete scheme for the elastic displacements and the dynamic liquid pressure is developed. The discrete scheme is based on the method of discrete singularities. A code to perform the numerical analysis is developed. Comprehensive benchmarking of the obtained results against other methods is done and good agreement is observed. The convergence of the proposed numerical method is demonstrated. One of the advantages of this new model is that the initial 3D problem is analytically reduced to a 1D integral equation. Moreover, it can handle the behaviour of the pressure in the vicinity of the nodes explicitly and the computational technique used has a quick convergence requiring a negligible amount of CPU time.
\end{abstract}


\end{frontmatter}


\section{Introduction}
\label{Introduction}
A great variety of storage containers are used for industrial as well as for many other purposes. In most cases these can be represented as compound shells of revolution. They could contain fuel, water, waste propellant or other liquids. For example water containers are used to collect, transport, treat, store and consume water.

A lot of publications have been devoted to study the fluid-elastic characteristics for shells which contain a liquid. In \citep{Ibrahim} the author considers them systematically from basic theory to advanced analytical and experimental results. In \citep{Chen_Hwang_Ko} authors develop the numerical method for the problem of natural vibrations for the fluid-filled elastic shells of revolution using the boundary element method (BEM). In \citep{Ventsel} authors describe the method of determining the normal modes of vibrations and natural frequencies of elastic shells of revolution with an arbitrary meridian, partially filled with a fluid using BEM and the finite element method (FEM).

In this paper the analytic mathematical models with corresponding discrete schemes for the elastic displacements and the dynamic liquid pressure are developed. The initial 3D problems is reduced to 2D by using the rotational symmetry of the shell. The equations of equilibrium and motion are obtained for an empty shell taking into account the variational principle of elasticity theory. The 2D problem for the dynamic liquid pressure is then simplified to a 1D boundary integral equation. The behaviour of the pressure in the vicinity of the shell nodes is taken into account analytically.

The shells are assumed to consist of spherical, cylindrical and conical surfaces (see Fig. 1). The liquid is considered to be ideal and incompressible.
\begin{figure}[h]
\centering\includegraphics[height=2.5in]{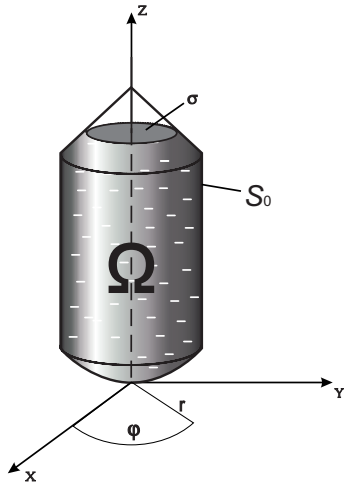}
\caption{An example of the considered shell, i.e. combination of spherical, cylindrical and conical surfaces} \label{shell_full}
\end{figure}

\section{Problem statement}
\label{ProblemStatement}
The total pressure $P(r,\varphi,z,t)$ in the domain $\Omega$ which is filled by the liquid can be considered as a superposition of the static pressure $P_{stat}(r,\varphi,z)$, corresponding to the liquid equilibrium, and the dynamic pressure $p(r,\varphi,z,t)$ which appears during the shell vibrations.

The elastic displacements of the shell and the dynamic pressure of the liquid on the shell need to be determined. Hence we consider equations of shell equilibrium for the elastic displacements $u(r,\varphi,z,t)$, $v(r,\varphi,z,t)$, $w(r,\varphi,z,t)$ and a Neumann problem for the dynamic pressure $p(r,\varphi,z,t)$ simultaneously:
\begin{equation} \label{main_system}
\left\{ \begin{array}{l}
\displaystyle L_r (u,v,w) = \rho h \frac{\partial^2 u(r,\varphi,z,t)}{\partial t^2} + q_r(r,\varphi,z,t); \\
\displaystyle L_\varphi (u,v,w) = \rho h \frac{\partial^2 v(r,\varphi,z,t)}{\partial t^2} + q_\varphi(r,\varphi,z,t); \\
\displaystyle L_z (u,v,w) = \rho h \frac{\partial^2 w(r,\varphi,z,t)}{\partial t^2} + p(r,\varphi,z,t) + q_z(r,\varphi,z,t); \\
\Delta p(r,\varphi,z,t) = 0.
\end{array} \right.
\end{equation}
where $L_r$, $L_\varphi$, $L_z$ are the differential operators  introduced in the variational principle of elasticity theory; $\rho$ -- the shell's material density, $h$ -- the shell thickness; $q_r(r,\varphi,z,t)$ and $q_\varphi(r,\varphi,z,t)$ and $q_z(r,\varphi,z,t)$ -– the spatial components of the driving force $\vec{q}$, which is known and axially symmetrical.

From now on we consider the compound shell as a set of parts $\Gamma_i$, where $\Gamma_1$ corresponds to the shell bottom, $i = \overline{1,N_{parts}}$.

Depending on how the shell is fastened one of the following boundary conditions for the elastic displacements is considered:
\begin{description}
  \item[(a)] the top and the bottom of the shell are fixed, hence the components of the generalized displacement $\vec{U}$ are equal to zero for $\Gamma_1$ and $\Gamma_{N_{parts}}$, i.e. Dirichlet boundary condition:
\begin{equation} \label{top_bottom_fixed}
\left. \vec{U} \right|_{\Gamma_1,\Gamma_{N_{parts}}} = 0
\end{equation}
  \item[(b)] the bottom of the shell is fixed and the top is free, hence the components the of generalized displacement $\vec{U}$ are equal to zero for $\Gamma_1$ and the components of generalized force $\vec{T}$ are equal to zero for $\Gamma_{N_{parts}}$, i.e. Neumann boundary condition:
\begin{equation} \label{bottom_fixed}
\left. \vec{U} \right|_{\Gamma_1} =0; \quad \left. \vec{T} \right|_{\Gamma_{N_{parts}}} = 0
\end{equation}
\end{description}

In both cases a) and b) the generalized displacement and the force should be continuous at the junctions $N_i$ between the different parts of the shell, $i = \overline{1,N_{parts}-1}$:
\begin{equation} \label{junctions}
\left. \vec{U} \right|_{N_i} = \left. \vec {U} \right|_{N_{i + 1}}; \quad \left. \vec {T} \right|_{N_i} = \left. \vec {T} \right|_{N_{i + 1}}.
\end{equation}

The system of equations (\ref{main_system}) is considered together with the boundary conditions for the elastic displacements (\ref{top_bottom_fixed}), (\ref{junctions}) or (\ref{bottom_fixed}), (\ref{junctions}) and with the following boundary condition for the dynamic pressure of the liquid:
\begin{equation}\label{boundary_condition}
\frac{\partial p}{\partial n} = \left\{ \begin{array}{l}
{ - \rho _l \left. {\left[ \displaystyle {\frac{\partial \vec {v_0}}{\partial t}\vec {n} +
\frac{\partial \vec {a}}{\partial t}\left( {\vec {s}\times \vec {n}}
\right) + \frac{\partial ^2w}{\partial t^2}} \right]} \right|_{S_0 } } \\
{ - \rho _l \left. {\left[ \displaystyle {\frac{\partial \vec {v_0}}{\partial t}\vec {n} +
\frac{\partial \vec {a}}{\partial t}\left( {\vec {s}\times \vec {n}}
\right) + \frac{\partial ^2f}{\partial t^2}} \right]} \right|_\sigma } \\
\end{array} \right.,
\end{equation}
where $\rho_l$ is the liquid density; $\vec{v_0}$ -- the velocity; $\vec{a}$ -- the angular acceleration; $\vec{s}$ -- the radius-vector; $\vec{n}$ -- the outer normal to the shell surface; function $f(r,\varphi,z,t)$ represents the free liquid surface; $S_0$ -- the shell surface and $\sigma$ -- the free surface of the liquid.

The free surface can be described in different ways \citep{Chen_Kuang_Jiao, Ventsel}. We consider vibrations for which the free surface stays planar, so the function $f(r,\varphi,z,t)$ is a function of the time only:
\begin{equation}
\label{FreeSurface}
f\left( t \right) = - \frac{1}{|\sigma|} \iint\limits_{S_0 } w dS_0.
\end{equation}
where $|\sigma|$ is a square of the free surface.

The initial conditions for problem (\ref{main_system}) in the presence of the driving force $\vec{q}$ have the following form:
\begin{equation}
\begin{array}{l}
u(r,\varphi,z,0) = v(r,\varphi,z,0) = w(r,\varphi,z,0) = f(r,\varphi,z,0) = 0; \\
u_t(r,\varphi,z,0) = v_t(r,\varphi,z,0) = w_t(r,\varphi,z,0) = f_t(r,\varphi,z,0) = 0; \\
p(r,\varphi,z,0) + \rho_l g f(r,\varphi,z,0) = 0,  \end{array}
\end{equation}
where $g$ is the gravitational acceleration constant.

To solve the boundary problem (\ref{main_system}-\ref{boundary_condition}), we represent the unknown functions in the following way:
\begin{equation}\label{series_by_t}
\begin{array}{l}
u(r,\varphi,z,t) = \sum \limits_{k=1}^\infty c_k(t) u_k(r,\varphi,z),\\
v(r,\varphi,z,t) = \sum \limits_{k=1}^\infty c_k(t) v_k(r,\varphi,z),\\
w(r,\varphi,z,t) = \sum \limits_{k=1}^\infty c_k(t) w_k(r,\varphi,z),\\
p(r,\varphi,z,t) = \sum \limits_{k=1}^\infty \ddot{c}_k(t) p_k(r,\varphi,z)
\end{array}
\end{equation}
and consider the derived problem for the functions $u_k(r,\varphi,z)$, $v_k(r,\varphi,z)$, $w_k(r,\varphi,z)$, $p_k(r,\varphi,z)$, $k = \overline{1,\infty}$. Further we will omit the $k$ index.

Profiting from the rotational symmetry, $u(r,\varphi,z,t)$, $v(r,\varphi,z,t)$, $w(r,\varphi,z,t)$ and $p(r,\varphi,z,t)$ in the form of the Fourier series:
\begin{equation} \label{Fourier_series}
\begin{array}{l}
u(r,\varphi,z) = u_0 (r,z) + \sum \limits_{m=1}^{\infty} [u_m^{(1)} (r,z) \cos m \varphi + u_m^{(2)} (r,z) \sin m \varphi], \\
v(r,\varphi,z) = v_0 (r,z) + \sum \limits_{m=1}^{\infty} [v_m^{(1)} (r,z) \cos m \varphi + v_m^{(2)} (r,z) \sin m \varphi], \\
w(r,\varphi,z) = w_0 (r,z) + \sum \limits_{m=1}^{\infty} [w_m^{(1)} (r,z) \cos m \varphi + w_m^{(2)} (r,z) \sin m \varphi], \\
p(r,\varphi,z) = p_0 (r,z) + \sum \limits_{m=1}^{\infty} [p_m^{(1)} (r,z) \cos m \varphi + p_m^{(2)} (r,z) \sin m \varphi].
\end{array}
\end{equation}

\section{Equlibrium Equations of the Elastic Shell}
\label{Equlibrium Equations}
The equilibrium equations mentioned in (\ref{main_system}) are derived from the basic variational principle of elasticity theory. These equations are obtained as Euler-Lagrange equations for the functional $J(u,v,w)$ which describes the difference between the potential energy of the thin elastic shell and the work performed by the external forces:
\begin{multline}
J(u,v,w) = \frac{1}{2} \int \limits_V \left[ \frac{\nu}{1 - 2 \nu} (e_{rr} + e_{\varphi \varphi} + e_{zz})^2 +
e^2_{rr} + e^2_{\varphi \varphi} + e^2_{zz} + \right. \\ \left. + \frac{1}{2} (e^2_{zr} + e^2_{z \varphi} + e^2_{r \varphi}) \right]
\frac{E dV}{1 + \nu} - \int \limits_V (Q_r u + Q_\varphi v + Q_z w) dV - \\ - \int \limits_S (q_r u + q_\varphi v + q_z w) d\sigma,
\end{multline}
where $Q_r$, $Q_\varphi$, $Q_z$ are the projections of the volume force $\vec{Q}$; $V$ is the domain occupied by the elastic shell; $S$ is the boundary surface; $E$ is the elasticity coefficient; $\nu$ is the Poisson's ratio; $e_{rr}$, $e_{\varphi \varphi}$, $e_{zz}$ -– the relative dilatations, $e_{zr}$, $e_{z \varphi}$, $e_{r \varphi}$ -– the shift angles.

Using the representations (\ref{series_by_t}-\ref{Fourier_series}), we obtain corresponding ordinary differential equations system in the domain $D = \Omega \bigcap xOz$ (see Fig. 2), which we solve by the method of generalized differential quadratures \citep{Kononenko_Strelnikova}.
\begin{figure}[h]
\centering\includegraphics[height=2.2in]{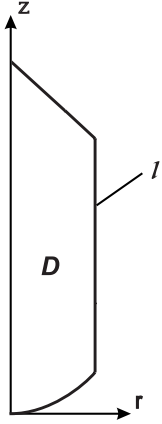}
\caption{2D-crossection of the shell shown on Fig. 1} \label{2D_shell}
\end{figure}

\section{Dynamic Pressure of Liquid to the Elastic Shell}
\label{Dynamic Pressure}
Using the representations (\ref{series_by_t}) and (\ref{Fourier_series}) the boundary problem (\ref{main_system}-\ref{boundary_condition}) for dynamic pressure $p(r,\varphi,z,t)$ is rewritten for the unknown functions $p_m^{(1,2)}, m = \overline{0,\infty}$ in domain $D$ (see Fig. 5):
\begin{equation} \label{SDE_op}
L_0 p_0 = 0, L_m p_m^{(1,2)} = 0,
\end{equation}
\begin{equation} \label{SDE_bc}
\frac{\partial p_0}{\partial n} = F_0, \frac{\partial p_m^{(1,2)}}{\partial n} = F_m^{(1,2)}.
\end{equation}
\noindent where $\displaystyle L_0 p_0 =  \frac {\partial^2 p_0}{\partial z^2}
+ \frac {\partial^2 p_0}{\partial r^2} + \frac{1}{r} \frac{\partial p_0}{\partial r}$, $\displaystyle L_m p_m^{(1,2)} = \frac {\partial^2
p_m^{(1,2)}}{\partial z^2} + \frac {\partial^2 p_m^{(1,2)}}{\partial r^2} + \frac{1}{r} \frac{\partial p_m^{(1,2)}}{\partial r} -
\frac{m^2 p_m^{(1,2)}}{r^2}$, $F_0, F_m^{(1,2)}$ are coefficients in the form of the Fourier series for the right-hand side of (\ref{boundary_condition}).
	
The fundamental solution of the equations (\ref{SDE_op}) is a function
\begin{equation} \label{f_m}
G_m (M, M_0) = \ln \frac{1}{\sqrt{|M-M_0|}} + H_m(M,M_0)
\end{equation}	
where $M(r,\varphi), M_0(r_0,\varphi_0) \in D$, $H_m(M, M_0)$ is a smooth function \citep{Kononenko}.
	
The solution of the boundary problem (\ref{SDE_op}-\ref{SDE_bc}) can be represented in terms of the single layer potential
\begin{equation} \label{initial}
p_m(r,z) = \int \limits_l g_m(M_0) G_m(M,M_0) dl, M \in D, M_0 \in l
\end{equation}
where the function $g_m (s_0)$ is a solution of the following Fredholm integral equation
\begin{equation} \label{fredholm}
g_m(s_0) = \frac{1}{\pi} \int \limits_l \left( \frac{\sin \theta}{2 r_0} \ln \frac{1}{|M'-M_0|} - Q_m(s_0,s') \right) g_m(s') ds' + \frac{1}{\pi} F_m(s_0),
\end{equation}
and the kernel $Q_m(s_0,s')$ of this equation can be represented in the following form:
\begin{equation} \label{kernel_Q}
Q_m(s_0,s') = \frac{\sin (\theta - \theta_0)}{|M'-M_0|} + \left. \left( \frac{\partial H_m}{\partial n} + \frac{\sin \theta}{2 r_0} \ln \frac{1}{|M'-M_1|} \right) \right|_{M \to M_0},
\end{equation}
where $\theta$ is the angle between the outer normal $n$ to the curve $l$ at the point $M_0$ and the positive $z$ axis direction,
$\theta_0$ is the angle between the outer normal $n_0$ to the line $M_0M'$ and the positive $z$ axis direction (see Fig.3).
\begin{figure}[h]
\centering\includegraphics[height=2.0in]{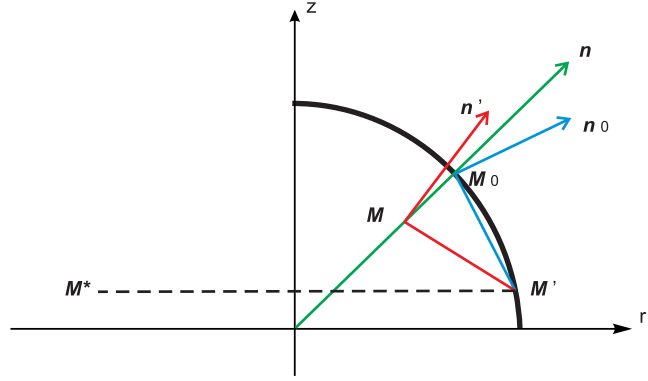}
\caption{Schematic view of the generatrix $l$}\label{fredholm_pic}
\end{figure}
									
By parameterizing and decomposing the curve $l$, the following system of boundary integral equations is derived from (\ref{fredholm}) for each $\Gamma_i$, $i = \overline{1,N-1}$:
\begin{equation} \label{BIE}
g^i_m(s_0) - \frac{1}{\pi} \int \limits_l \left( \frac{\sin \theta}{2 r_0} \ln \frac{1}{|M'-M_0|} - Q^i_m(s_0,s') \right) g^i_m(s') ds' = \frac{1}{\pi} F^i_m(s_0),
\end{equation}
as defined in Section 2.

Using special quadrature formulas of the interpolative type with nodes which are zeros of the Chebychev polynomials of second kind \citep{Gandel_Kononenko, Lifanov} the following system of linear algebraic equations is obtained for each $\Gamma_i$, $i = \overline{1,N-1}$:
\begin{multline}\label{SLAE}
f^i_m(t^n_{0j}) \sqrt{1-(t^n_{0j})^2} - \sum \limits_{k=1}^{n-1} \left(\displaystyle \frac{\sin \theta}{2r_0} \ln \frac{1}{|t^n_{0k} - t^n_{0j}|} - Q^i_m(t^n_{0k},t^n_{0j}) \right) f^i_m(t^n_{0k}) = F^i_m(t^n_{0j}),
\end{multline}
where $j = \overline{1,n-1}$, $n$ is the number of discretisation points along $\Gamma_i$ and
\begin{equation}\label{f_m_dm}
\displaystyle f^i_m(t) = \frac{g^i_m(t)}{\sqrt{1-t^2}}.
\end{equation}

We can determine the behaviour of the solution of (\ref{BIE}) in the vicinity of the shell nodes by considering the angle $\alpha$ between $\Gamma_i$ and $\Gamma_{i+1}$, $i = \overline{1,N-1}$ as an angle between corresponding tangents in the mutual node (see Fig. 4).
\begin{figure}[!h]
\centerline{\includegraphics[height=1.5in]{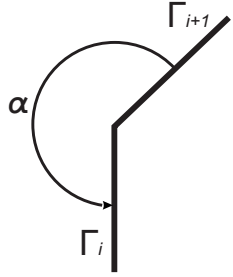}}
\caption{Vicinity of the shell node}
\end{figure}

For $0 < \alpha \leq \pi$ the solution of integral equation is a smooth function, $\pi < \alpha < 2 \pi$ the solution has an singularity which has an order of $1/r^{1-\frac{\pi}{\alpha}}$, for $\alpha = 2 \pi$ the solution has a hypersingularity and this case corresponds to the internal edge inside the shell. This solution behaviour is taken into account in the form of $f^i_m (t)$ function in (\ref{f_m_dm}).

\section{Numerical Results}
\label{Numerical Results}
We consider a shell composed of cylindrical, conical and spherical surfaces with a flat bottom partially filled with the liquid. The geometry of the tank is shown in Fig. 5. The radiuses of the spherical and cylindrical parts are 1 m, the height of spherical part is 1 m, the height of cylindrical part is 2 m, the height of conical part is 2 m, the level of filling is 4 m, the shell thickness $h$ is 0.01 m.

The relevant mechanical parameters of the considered shell are: Young modulus $E$ = 2·105 MPa, Poisson ratio $\nu$ = 0.3, yield point $T$ = 320 MPa, shell material density $\rho$ = 7800 kg/m3, liquid density $\rho_0$ = 1000 kg/m3, fixed bottom, the pulse load $q(t)$: $q_0$ = 0.1 MPa, $\tau$ = 14.2E-6 s, $q(t) = q_0 \exp(- t / \tau)$.
\begin{figure}[h]
\centering\includegraphics[height=2.5in]{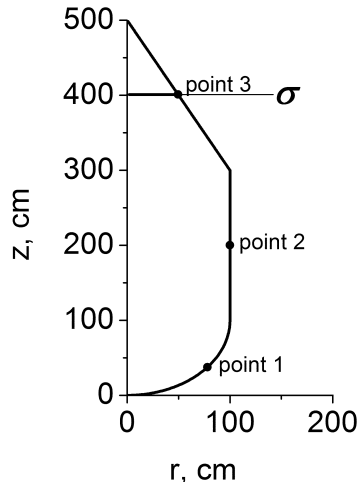}
\caption{2D-crossection of the considered shell}\label{Geometry}
\end{figure}

In Fig. 6 and 7 we show the results of benchmarking for both the normal displacements of the shell as calculated by our approaching (solid lines) and BEM in ANSYS (dash lines) in point 1 (40,80), point 2 (100, 200), point 3 (50, 400) that are shown in Fig. 5.
\begin{figure}[!h]
\centerline{\includegraphics[height=2.5in]{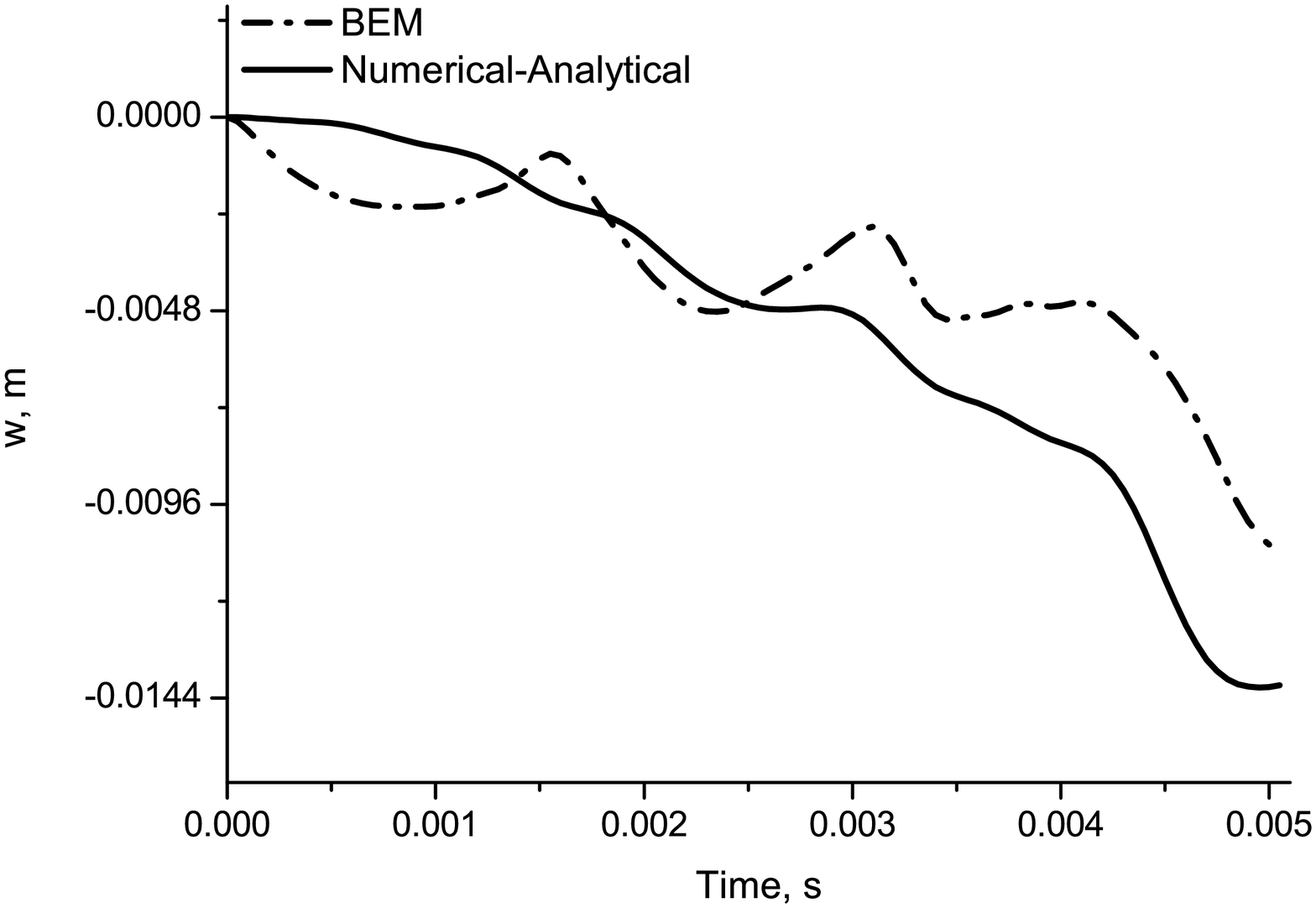} \includegraphics[height=2.5in]{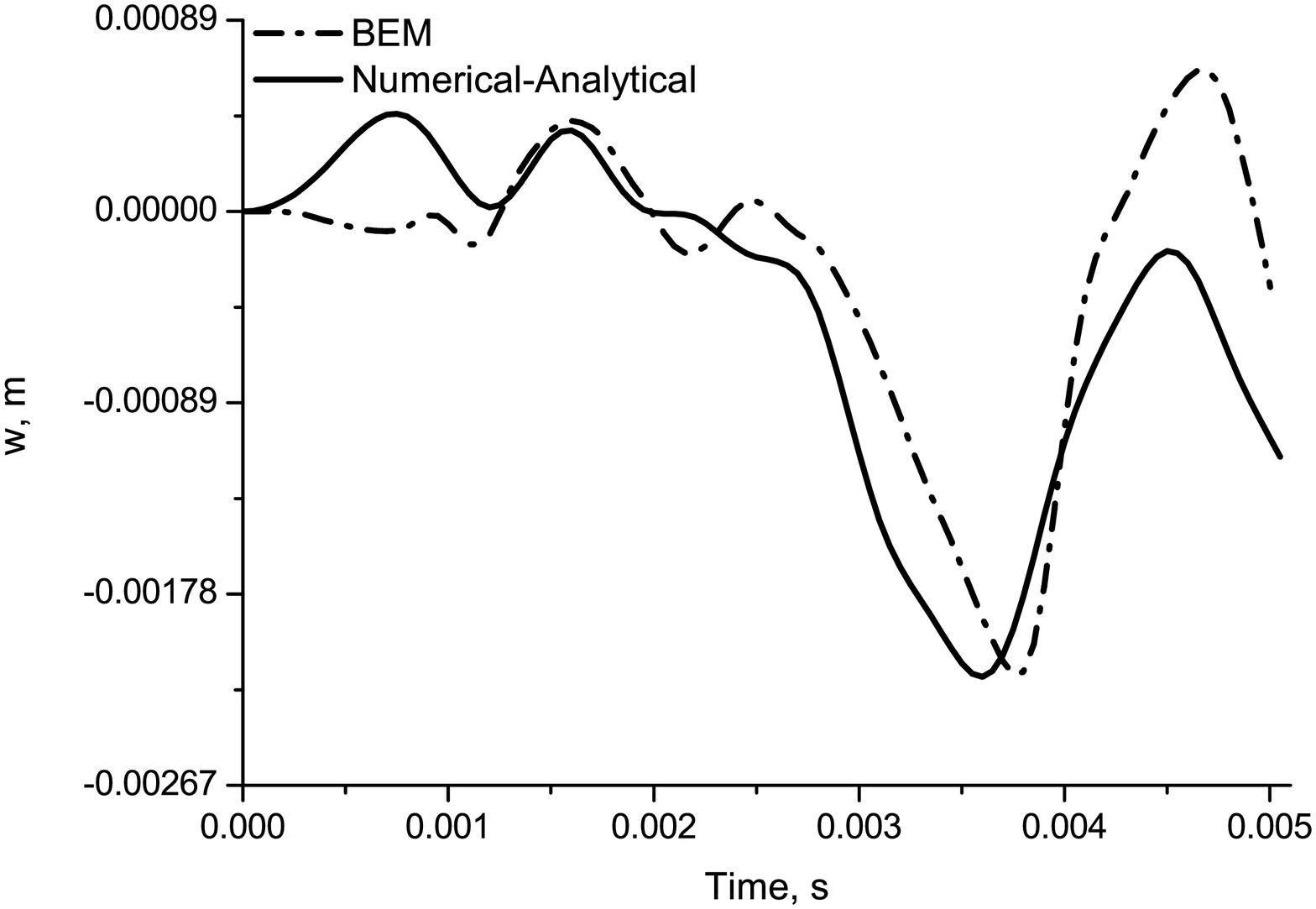}}
\caption{The normal displacement in points 1 (left) and 2 (right)}
\end{figure}
\begin{figure}[!h]
\centerline{\includegraphics[height=2.5in]{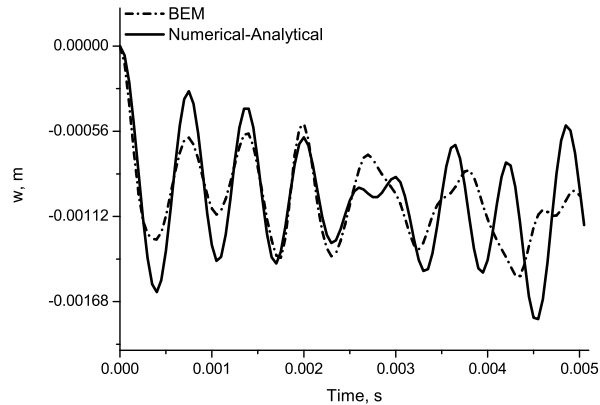}}
\caption{The normal displacement in point 3}
\end{figure}

In Fig. 8 we show the results for the liquid pressure in these three points
\begin{figure}[!h]
\centerline{\includegraphics[height=2.5in]{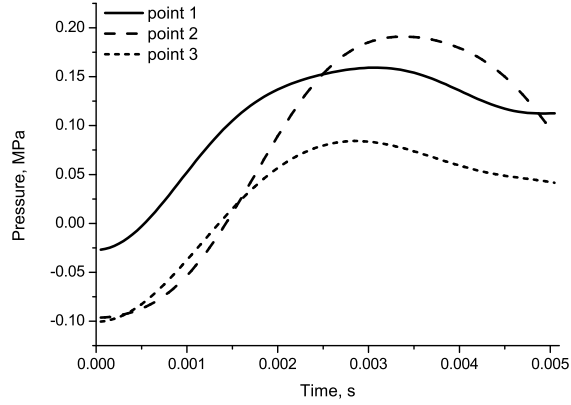}}
\caption{The dynamic liquid pressure}
\end{figure}

The extremely fast convergence of our method is illustrated in Fig. 9 on different shell parts: $\varepsilon_1$ -- relative convergence on spherical part, $\varepsilon_2$ -- relative convergence on cylindrical part, $\varepsilon_3$ -- relative convergence on conical part.
\begin{figure}[htbp]
\centerline{\includegraphics[height=2.2in]{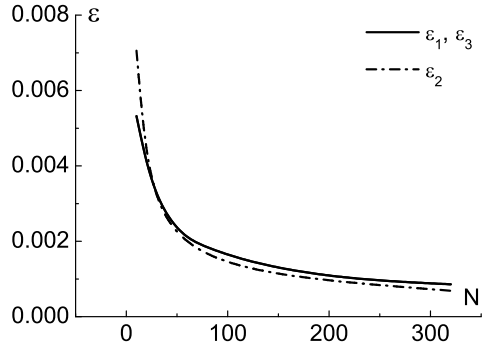}}
\caption{Convergence of the proposed method} \label{conv_v_test}
\end{figure}

These results of BEM (\cite{Brebbia}) and method proposed in this paper are in a good agreement. While providing the same level of accuracy, our method has fast convergence and requires negligible amount of CPU time.

\section{Conclusions}
\label{Conclusions}
The method proposed in this paper can be applied to solve the following problems:
\begin{description}
  \item[(a)] static problems: right-hand sides in (\ref{main_system}) equal to zero;
  \item[(b)] free vibration problems of empty shells: $P = 0$, $q_r = q_\varphi = q_z = 0$;
  \item[(c)] free fluid-elastic vibrations of partially filled shells: $P \neq 0$,\\ $q_r = q_\varphi = q_z = 0$;
  \item[(d)] forced vibration problems of empty shells: $P = 0$, $q_r, q_\varphi, q_z \neq 0$;
  \item[(e)] forced fluid-elastic vibrations of partially filled shells: $P \neq 0$,\\ $q_r, q_\varphi, q_z \neq 0$.
\end{description}

A new mathematical model of hydroelastic vibrations of partially filled shells is developed. Elastical displacements are determined by solving equilibrium equations. The 3D Neumann problem for dynamic pressure is reduced to the corresponding 2D one by using the rotational symmetry of the shell. This 2D problem is then simplified to a 1D integral equation for pressure difference.

A corresponding discrete mathematical model is built based on the Chebyshev polynomials. Simulation codes are developed to determine the dynamic pressure acting on the shell and the free liquid surfaces.  The ANSYS package is used to compare and validate the results.

The advantage of this new model is that it can handle the behaviour of the pressure in the vicinity of the nodes analytically by reducing the initial 3D problem to a 1D integral equation. Moreover the computational technique has a quick convergence and requires a negligible amount of CPU time.

The research presented in this paper has been done within the framework of the Ukrainian Governments R\&D projects.

The author is grateful to Prof. E. Strelnikova for her valuable guidance and support of this work and to Prof. Yu. V. Gandel for stimulating discussions.

\bibliographystyle{arxivarticle-harv}
\bibliography{Kononenko_arxiv}







\end{document}